\documentclass[aps,physrev,reprint,twocolumn,superscriptaddress]{revtex4-2}

\usepackage[T1]{fontenc}
\usepackage[utf8]{inputenc}
\usepackage{amsmath,amssymb,amsthm,mathtools,bm}
\usepackage{physics}
\usepackage{microtype}
\usepackage{enumitem}
\usepackage{orcidlink}
\usepackage{hyperref}

\newlength{\ketketwidth}
\newlength{\ketwidth}

\newcommand{\kettstylesep}[3]{
    \settowidth{\ketwidth}{$#2\left|#1\right\rangle$}
    \settowidth{\ketketwidth}{$#2\left.\left|#1\right\rangle\right\rangle$}
    \left|#1\right\rangle#3\hspace{\ketwidth}\hspace{-\ketketwidth}
}
\newcommand{\kett}[1]{
    \left.\mathchoice
        {\kettstylesep{#1}{\displaystyle}{\hspace{0.3em}}}
        {\kettstylesep{#1}{\textstyle}{\hspace{0.3em}}}
        {\kettstylesep{#1}{\scriptstyle}{\hspace{0.3em}}}
        {\kettstylesep{#1}{\scriptscriptstyle}{\hspace{0.25em}}}
    \right\rangle
}

\newcommand{\bbrastylesep}[3]{
    \settowidth{\ketwidth}{$#2\left\langle#1\right|$}
    \settowidth{\ketketwidth}{$#2\left\langle\left\langle#1\right|\right.$}
    #3\hspace{\ketwidth}\hspace{-\ketketwidth}\left\langle#1\right|
}
\newcommand{\bbra}[1]{
    \left\langle\mathchoice
        {\bbrastylesep{#1}{\displaystyle}{\hspace{0.3em}}}
        {\bbrastylesep{#1}{\textstyle}{\hspace{0.3em}}}
        {\bbrastylesep{#1}{\scriptstyle}{\hspace{0.3em}}}
        {\bbrastylesep{#1}{\scriptscriptstyle}{\hspace{0.25em}}}
    \right.
}

\newcommand{\kettbbra}[1]{\kett{#1}\hspace{-0.3em}\bbra{#1}}

\allowdisplaybreaks

\newtheorem{definition}{Definition}
\newtheorem{proposition}{Proposition}
\newtheorem{theorem}{Theorem}
\newtheorem*{remark}{Remark}
\newtheorem{corollary}{Corollary}
\newtheorem{lemma}{Lemma}

\newcommand{\Hil}{\mathcal{H}}
\newcommand{\id}{\operatorname{id}}
\newcommand{\one}{\mathbb{I}}
\newcommand{\eqdef}{\mathrel{:=}}
\newcommand{\supp}{\operatorname{supp}}

\newcommand{\cU}{\mathcal{U}}

\newcommand{\cD}{\mathcal{D}}

\newcommand{\cM}{\mathcal{M}}
\newcommand{\cE}{\mathcal{E}}

\newcommand{\tp}{\textsf{T}}

\begin{document}

\title{Quantum Belief Characterization:
A Multi-Shot Notion of Equivalence }

\author{Mingxuan Liu\orcidlink{0009-0009-4361-2641}}
\email{liu.mingxuan@u.nus.edu}
\affiliation{Centre for Quantum Technologies, National University of Singapore, Singapore}

\author{Pavel Sekatski\orcidlink{0000-0001-8455-020X}}
\affiliation{Department of Applied Physics, University of Geneva, Switzerland}

\author{Valerio Scarani\orcidlink{0000-0001-5594-5616}}
\affiliation{Centre for Quantum Technologies, National University of Singapore, Singapore}
\affiliation{Department of Physics, National University of Singapore, Singapore}

\date{\today}

\begin{abstract}
A recent formulation of quantum belief showed that quantum retrodiction depends not only on the reduced density operator of the system of interest, but also on correlations with a hidden reference system, which leads to the notion of quantum belief equivalence. In this paper we formulate a multi-shot problem, where the extended posterior obtained after one update is used again as the prior for the next update. We first define a formal notion of $n$-shot equivalence and derive several propositions for arbitrary extended beliefs. We then argue that the agent must also specify why the data obtained in different rounds concern the same underlying object. Through the quantum de Finetti representation, this narrows the one-copy belief to a special class: candidate-state ensemble belief. In that sector, we show that two beliefs are $n$-shot equivalent if and only if their $(n+1)$-th moments are equal. We also give a physical interpretation of this moment condition and illustrate it through the equivalence between Haar priors and projective-design priors at bounded update depth.
\end{abstract}

\maketitle

\section{Introduction}

A quantum belief represents one's knowledge about a quantum system. As such, quantum beliefs are used in various quantum Bayesian inference tasks: in finite-data regimes, they provide optimal strategies for quantum state tomography~\cite{BQST2010,ABQST2012,Granade_2017} and quantum metrology~\cite{Helstrom,BaoMol20a,BaoMol20b}; they also play a central role in quantum retrodiction~\cite{ABL64,Barnett2000,Fuchs03,Tsa-PRL09,GJM13,LS13,GueWis15,BusSca21,AwBS,LiuBaiScarani2025unifying}, inferring the state of a quantum system in the past. Quantum beliefs also serve as the basis for the Bayesian interpretation of quantum theory, known as QBism~\cite{fuchs2010qbism}, which aims at advancing the understanding of enigmatic problems in quantum physics~\cite{CavesFuchsSchack2002QB,CavesFuchsSchack2002dF, FuchsSchack2013RMP, FuchsMerminSchack2014, debrota2020respecting}.

Reduced density operators were long taken to provide a complete description of quantum beliefs. If $S$ is part of a larger system $SR$, it is the reduced density operator that predicts the statistics of outcomes for any local measurement on $S$, regardless of which purification, ensemble decomposition, or inaccessible reference system realizes it~\cite{d1976conceptual,HJW93,nielsen2010qcqi}. However, Ref.~\cite{liu2026properimproper} showed that two beliefs with the same reduced density operator can induce different retrodictive updates when they arise from different correlations with a hidden reference system. Thus, for retrodiction the relevant prior is the extended belief on $SR$, and not just the reduced density operator on $S$. 
More concretely, in the prior-extended Petz map framework~\cite{petz1986,petz1988,liu2026properimproper},  the output of a channel acting on $S$ is pulled back to an updated belief on $SR$, and the corresponding {\it updated} belief on $S$ is obtained by tracing out $R$.
The resulting retrodiction map depends on the full extended prior, not only on its marginal.

This leads to the notion of equivalence between beliefs: two extended beliefs are equivalent if they induce the same retrodiction map for every channel. This notion is intrinsically single-shot. Operationally, it asks whether two extended priors always yield the same updated belief on the system of interest ($S$), for any admissible channel and evidence. In this paper, we extend the notion to iterative updates, in which the extended posterior obtained in one round serves as the prior for the next. This procedure is mathematically well-defined for arbitrary extended beliefs and we analyze it below, but its physical interpretation is not automatic and calls for some care.

In ordinary Bayesian inference, repeated data are informative because they concern the same underlying object, parameter, or source. A doctor may repeatedly update their belief about whether a patient is infected because the tests are understood to be effectively non-disturbing and to concern the same patient. If the test results instead came from unrelated patients, iterating the same update would no longer describe learning about any one patient. The issue is sharper in quantum theory. In general, one cannot repeatedly measure an unknown system while leaving its state unchanged, because quantum measurements have back-action; nor can one produce additional copies of an unknown state by cloning~\cite{wootters1982nocloning,DIEKS1982271} or broadcasting~\cite{BCFJS96nobroadcasting}. Consequently, a sequence of quantum evidence cannot generally be interpreted as repeated access to one unchanged system \footnote{An agent can nevertheless infer the past state of a quantum system from the outcomes of a sequence of measurements performed successively on that system. The task, known as quantum state smoothing, is tackled by incorporating the full measurement record, together with measurement back-action, into a single retrodictive inference~\cite{LiuBaiScarani2025unifying}. This differs from the iterative update considered here.}. We argue that the iteration can be understood naturally under a conditionally i.i.d. source model that relates the different systems from which the evidence is obtained.

The plan of the paper is as follows. After introducing preliminary notions (Section \ref{sec: formal}), we first present a formal construction of iterative quantum belief updates (Section \ref{sec:multi}). This allows us to introduce the notion of multi-shot equivalence for arbitrary beliefs and give conditions for the equivalence to hold. Then, via the de Finetti representation~\cite{Hudson1976,CavesFuchsSchack2002dF}, we identify \textit{candidate-state ensemble beliefs} as the natural single-copy analogue of an agent's belief that a source prepares an exchangeable sequence of systems (Section \ref{sec:candidate-ensemble-belief}). We prove our main result (Theorem \ref{thm:proper}): two candidate-state ensemble beliefs are $n$-shot equivalent if and only if their $(n+1)$-th moments coincide. We also provide an operational meaning for this result, and illustrate it with an example based on quantum state designs.

\section{Preliminary}
\label{sec: formal}

\subsection{Classical belief update}
In classical probability theory, the well-known Bayes' rule~\cite{bayes1763essay} gives the canonical prescription for updating prior belief, while Jeffrey's rule~\cite{Jeffrey} extends it to the case of ``soft evidence’’.  Concretely, let  $X$ be a random variable with prior distribution $p(x)$, and $\mathcal{P}(y|x)$ a stochastic process inducing a new variable $Y$. Suppose that an agent receives evidence that the marginal distribution of $Y$ is $q(y)$. Jeffrey's rule provides a prescription to update the agent's prior belief about $X$ based on this evidence
\begin{equation}\label{eq: Jeffrey}
p' = \mathcal{J}_{\mathcal{P},q}(p), \quad \text{with} \quad p'(x) = p(x)\sum_y\frac{q(y) \mathcal{P}(y|x)}{\mathcal{P}_*p(y)}.
\end{equation}
and $\mathcal{P}_* p(y)= \sum_x p(x) \mathcal{P}(y|x)$. The usual Bayesian update $p'(x)=p(x|y')$ is recovered for ``hard evidence'' $q(y)=\delta_{y,y'}$. It is also interpreted as a way of retrodiction when used for one-shot problems, e.g., estimating the past state of a system that undergoes a  stochastic process based on presently available information.
In scenarios where evidence $q_i$ associated to maps $\mathcal{P}_i$, which are obtained from the same object, is received sequentially the agent's belief can be updated iteratively by applying the maps $ \mathcal{J}_{\mathcal{P}_i,q_i}$ one after the other. Notably, these maps do not always commute, hence the final belief may depend on the order in which channel-evidence pair $(\mathcal{P}_i,q_i)$ is obtained~\cite{diaconis1982updating}. The intuitive origin of this non-commutativity is that soft evidence does not represent objective data, but is only interpreted meaningfully relative to the agent’s current state of belief.

For illustration, consider the two following examples. For the first example, mentioned in the introduction, $p(x)$ describes the doctor's prior belief in a patient being infected with various diseases, based e.g. on prevalence in the population, while $q(y)$ is the soft evidence coming from new test results. Since the testing procedure may be regarded as effectively nondestructive, the updated distribution $p'(x)$ can be understood as doctor's current belief about patient's diagnosis, in accordance with Jeffrey's original interpretation. For the second example, let $p(x)$ describe the experimenter's belief in the conformational state of a molecule, and $q(y)$ describe noisy fluorescence signal. Here, the stochastic map $\mathcal{P}: X\to Y$ describing the fluorescence measurement is intrinsically disturbing of molecule's state. Hence, the updated distribution $p'(x)$ can only be interpreted as a revised belief about the molecule's state in the past, but not as a belief about its present state. 

The two examples are even further apart  when considering iterative belief updating. The first example naturally generalizes to a sequential scenario -- the doctor performs new tests and updates the belief. In the second example, the molecule's initial state was irreversibly altered, and can not be probed again. No new evidence from the same state of the molecule would possibly be attained. As a consequence, iterative updating is not possible unless there are many copies of the original molecule, and the experimenter's belief is not merely about the state of the concrete molecule but is associated with their preparation procedure, i.e., about a {\it source} preparing copies of a system.

\subsection{One-shot quantum belief update}

As we argued, retrodiction generally concerns statements about systems that existed in the past and may no longer be accessible for direct verification. In classical probability theory this limitation is not fundamental. Indeed, one may imagine that, prior to the measurement, the state of the system $X$ was perfectly copied to an auxiliary system $X'$, so that the joint distribution satisfies $p(x,x')=p(x) \delta_{x,x'}$. Consequently, the revised belief $p'(x)$ about the past can always be interpreted operationally as a bet on the state of the untouched copy $X'$.

The situation is more subtle in quantum theory, where unknown states can neither be cloned nor read out in a single shot. Nevertheless, a quantum retrodiction map can be introduced formally, and several proposals have been explored~\cite{ozawa1997quantum,fuchs2001quantum,parzygnat2022non,Tsa-PRA22,PF23,PB23,SS23,Scandi25QFIPetz,Ge2025MCP}. The question of this paper can be posed for any retrodictive update
rule,  but we analyze it for the Petz recovery map~\cite{petz1986,petz1988} as the quantum analogue of Bayes' and Jeffrey's rules given its distinguished properties and wide acknowledgment~\cite{PB23,Ge2025MCP,Scandi25QFIPetz,murk2026connectingquantumtomographyquantum}.

Following Ref.~\cite{liu2026properimproper}, a \emph{quantum belief} is a density operator $\beta\in \cD(\Hil_S\otimes\Hil_R)$ on both the system under study $S$ and a reference $R$, which is hidden, but whose correlations with $S$ may affect the update. The corresponding belief about the reduced density operator of system $S$ is denoted by $\beta_S=\Tr_R[\beta]$. Given a channel $\cE:S\to T$ and the output state $\sigma\in \cD(\Hil_T)$, the update on the belief about the reduced density operator of system $S$ is determined by the channel-evidence pair $(\cE,\sigma)$ as
\begin{align}
    \beta'_S=\Tr_R[\beta']=\Tr_R\bigl[\cU_{\cE,\sigma}(\beta)\bigr]\,,
\end{align}
where we have defined the prior-extended Petz update
\begin{equation}\label{eq:extended-update}
\cU_{\cE,\sigma}(\beta)
= \sqrt{\beta}\bigl(\cE^\dagger\!\left(
\cE(\beta_S)^{-1/2}\sigma\,\cE(\beta_S)^{-1/2}
\right)\otimes \one_R\bigr)\sqrt{\beta}\,.
\end{equation}
In particular, it is worth noting that for classical beliefs, channels, and output states, the
prior-extended Petz update reduces to Jeffrey's rule; see
Appendix~\ref{app:classical-limit}.

The inverses in Eqs~\eqref{eq: Jeffrey} and \eqref{eq:extended-update} are defined on the relevant support. For the case where $\cE(\beta_S)$ is not full rank in $T$, we need a restriction mirroring the fact that some evidence can be incompatible with the prior belief.

\begin{definition}[Admissible update]
Let $\beta$ be an extended prior belief on $SR$, and let $\beta_S=\Tr_R[\beta]$.
A channel-evidence pair $(\mathcal E,\sigma)$ is called \emph{admissible for $\beta$} if
\begin{equation}
\supp(\sigma)\subseteq \supp(\cE(\beta_S)).
\end{equation}
Equivalently, the prior-extended Petz update \eqref{eq:extended-update}
is a normalized state: indeed, $\Tr[\cU_{\mathcal E,\sigma}(\beta)]=\Tr[\cE(\beta_S)\left(
\cE(\beta_S)^{-1/2}\sigma\,\cE(\beta_S)^{-1/2}
\right)]=\Tr[\Pi_{\supp(\cE(\beta_S))}\sigma]$.
\end{definition}

\section{Multi-shot notion of equivalence}
\label{sec:multi}

We now introduce iterative updates of quantum beliefs. In this section we focus on the formal iteration of update maps, the notion of multi-shot equivalent beliefs and its properties. The interpretation is postponed to the next section. As we have seen, even in the classical case, the interpretation of iterative updating sometimes requires a source model. In the case of quantum beliefs this requirement is even sharper and creates tension with the no-cloning theorem in general.  Without such a model, the mathematical formalism unfolded in this section is well defined, but the iteration mechanism may lack a physical meaning.

\subsection{Iterative quantum belief update}

Consider an update sequence of length $n$, a list
\begin{equation}
\mathbf{u}=\bigl((\cE_1,\sigma_1),\ldots,(\cE_n,\sigma_n)\bigr),
\end{equation}
where each $\cE_t:S\to T_t$ is a channel and each $\sigma_t\in \cD(\Hil_{T_t})$ is the corresponding evidence state. Starting from $\beta^{(0)}\eqdef \beta$, we define recursively
\begin{equation}
\beta^{(t)}(\mathbf{u})
\eqdef
\cU_{\cE_t,\sigma_t}\bigl(\beta^{(t-1)}(\mathbf{u})\bigr).
\end{equation}
Thus $\Tr_R\left[\beta^{(t)}(\mathbf{u})\right]$ is the updated belief about the reduced density operator of system $S$ after the first $t$ updates in the sequence. The sequence $\mathbf{u}$ is called \emph{admissible for $\beta$} if, for each $t=1,\ldots,n$ the pair
$(\mathcal E_t,\sigma_t)$ is admissible for the updated prior
$\beta^{(t-1)}(\mathbf u)$.

The Petz recovery maps do not commute in general, and hence the final belief $\beta^{(n)}$ can depend on the order in which evidence $(\cE_i,\sigma_i)$ is presented, just like for Jeffrey's rule. Nevertheless, in section~\ref{subsec:interpretation} we will see that commutativity is restored in the operationally relevant case of candidate-state ensemble beliefs and hard evidence. 

\subsection{Definition and structure}
\label{subsec:equivalence_condition}
We are now ready to define the notion of multi-shot equivalence of quantum beliefs.

\begin{definition}[\texorpdfstring{$n$}{n}-shot equivalence]
\label{def:nshot}
For two beliefs
$\beta\in\mathcal D(\mathcal H_S\otimes\mathcal H_{R_\beta})$ and
$\gamma\in\mathcal D(\mathcal H_S\otimes\mathcal H_{R_\gamma})$, we say that $\beta$ and $\gamma$ are \emph{$n$-shot equivalent}, written $\beta\sim_n\gamma$, if for any update sequence $\mathbf{u}$ of length $n$ admissible for both $\beta$ and $\gamma$ one has
\begin{equation}
\Tr_{R_\beta}\left[\beta^{(n)}(\mathbf{u})\right]=\Tr_{R_\gamma}\left[\gamma^{(n)}(\mathbf{u})\right].
\end{equation}
The two beliefs are \emph{iteratively equivalent} if $\beta\sim_n\gamma$ for any $n\geq 1$.
\end{definition}

\begin{figure}
    \centering
    \includegraphics[width=1\linewidth]{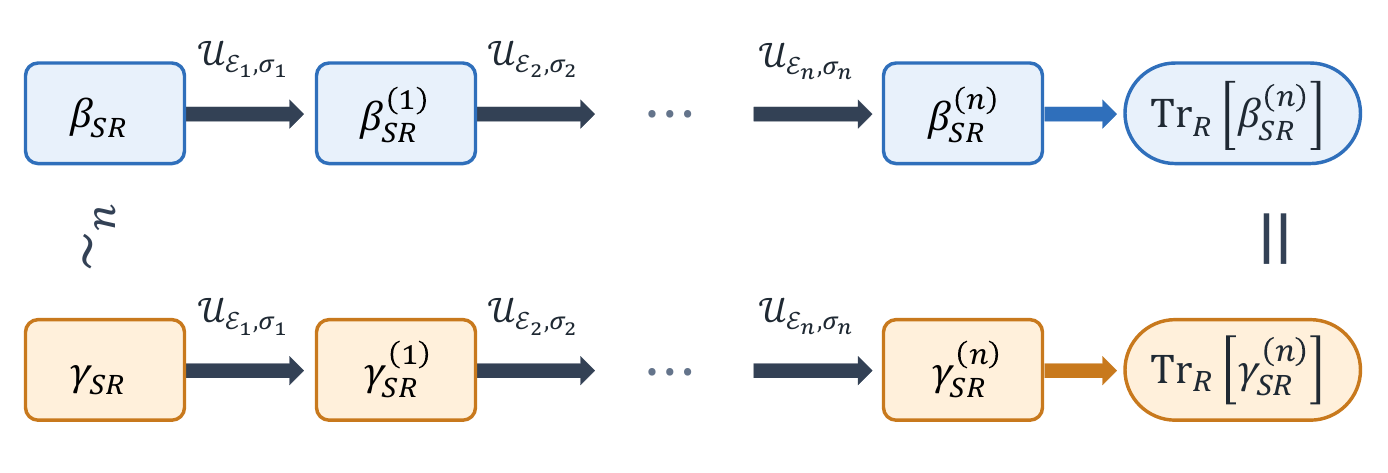}
    \caption{Multi-shot equivalence of quantum beliefs.
Starting from two extended beliefs $\beta$ and $\gamma$ on $SR$, the same admissible update sequence is applied. The beliefs are $n$-shot equivalent, $\beta\sim_n\gamma$, when the final posterior marginals on $S$ agree for any such sequence of length $n$.}
    \label{fig:nshot-equivalence}
\end{figure}

Figure~\ref{fig:nshot-equivalence} illustrates this definition: the relation $\beta\sim_n\gamma$ holds when applying the same admissible update sequence to the two extended beliefs always yields the same final posterior marginal on system $S$.
The notion is strictly stronger than single-shot equivalence because it requires the equality of the marginals of final belief after arbitrary sequences of updates rather than after only one.

\begin{lemma}[Idle update]
\label{lem:idle}
Let $\tau$ be any full-rank state on some output system $T$, and define the replacement channel
\begin{equation}
\Lambda(\rho)=\Tr[\rho]\,\tau.
\end{equation}
Then for any belief $\omega$ one has
\begin{equation}
\cU_{\Lambda,\tau}(\omega)=\omega.
\end{equation}
\end{lemma}

\begin{proof}
Since $\Lambda(\omega_S)=\tau$, $\Lambda^{\dagger}(\Lambda(\omega_S)^{-1/2}\tau\Lambda(\omega_S)^{-1/2})=\Lambda^{\dagger}(\tau^{-1/2}\tau\tau^{-1/2})=
\Lambda^{\dagger}(\one_T)=\one_S$.
Substituting into Eq.~\eqref{eq:extended-update} gives $\cU_{\Lambda,\tau}(\omega)=\omega$.
\end{proof}

\begin{proposition}[Padding]
\label{prop:padding}
If $\beta\sim_n\gamma$ and $m\le n$, then $\beta\sim_m\gamma$.
\end{proposition}

\begin{proof}
By definition, $\beta\sim_n\gamma$ means equivalence for all updates. Given any update sequence of length $m$, one can append $n-m$ idle updates of the form given in Lemma~\ref{lem:idle}, which do not change the belief. Equality after $n$ updates therefore implies equality after the first $m$ updates.
\end{proof}

The new definition immediately yields a recursive reduction.

\begin{proposition}[Recursive characterization]
\label{prop:recursive}
For $n\ge 2$, two beliefs $\beta$ and $\gamma$ satisfy $\beta\sim_n\gamma$ if and only if for any first update $(\cE,\sigma)$ that is admissible for both $\beta$ and $\gamma$ one has
\begin{equation}
\label{eq:recursive-characterization}
\cU_{\cE,\sigma}(\beta)\sim_{n-1}\cU_{\cE,\sigma}(\gamma).
\end{equation}
\end{proposition}

\begin{proof}
By definition, $\beta\sim_n\gamma$ means equality of the marginal beliefs obtained after the first update and then $n-1$ further updates, for any sequence of length $n$. This is exactly the condition that the once-updated extended priors be $(n-1)$-shot equivalent for any possible first update.
\end{proof}

Proposition~\ref{prop:recursive} is the general formulation of multi-shot equivalence. It reduces the $n$-shot problem to the single-shot problem applied to extended priors after $n-1$ updates. For single-shot update, a necessary and sufficient condition for equivalent beliefs is proved in \cite{liu2026properimproper}. One may apply Proposition~\ref{prop:recursive} to it and derive the $n$-shot equivalence condition for $n\ge2$, while a closed-form is not given in this paper.

\begin{proposition}[Sufficient conditions for iterative equivalence]
\label{prop:iterative-sufficient}
Let $\beta$ be a belief on $SR$. Let $A$ be an auxiliary system, let $\eta$ be a quantum state on $A$, and let
$V:\Hil_R\otimes\Hil_A\to \Hil_{\widetilde R}$
be an isometry. Define the channel
\begin{equation}
\label{eq:rev-channel-form}
\mathcal P(\rho)\eqdef V(\rho\otimes \eta)V^\dagger .
\end{equation}
Then $\beta$ and $(\id_S\otimes \mathcal P)(\beta)$ are iteratively equivalent.
\end{proposition}
In finite dimension, it is to say that quantum belief remains iteratively equivalent after any reversible channel acts on the reference system.
\begin{proof}
For any prior belief $\beta$ and
\begin{equation}
\tilde\beta\eqdef (\id_S\otimes \mathcal P)(\beta)
=
(\one_S\otimes V)(\beta\otimes \eta)(\one_S\otimes V^\dagger),
\end{equation}
they share the same marginal on system $S$,
\begin{equation}
\Tr_{\widetilde R}\bigl[\tilde\beta\bigr]
=
\Tr_{RA}\bigl[\beta\otimes \eta\bigr]
=
\Tr_R\bigl[\beta\bigr].
\end{equation}
For an admissible channel-evidence pair $(\mathcal E,\sigma)$, denoting
\begin{equation}
L\eqdef
\mathcal E^\dagger\!\left(
\mathcal E(\Tr_R[\beta])^{-1/2}\,
\sigma\,
\mathcal E(\Tr_R[\beta])^{-1/2}
\right),
\end{equation}
we have
\begin{align}
&\mathcal U_{\mathcal E,\sigma}(\tilde\beta ) =
\sqrt{\tilde\beta }\,(L\otimes \one_{\widetilde R})\,\sqrt{\tilde\beta } \nonumber\\
=&
(\one_S\otimes V)\bigl(\sqrt{\beta }\otimes \sqrt{\eta}\bigr)
(L\otimes \one_R\otimes \one_A)
\bigl(\sqrt{\beta}\otimes \sqrt{\eta}\bigr)
(\one_S\otimes V^\dagger) \nonumber\\
=&
(\one_S\otimes V)\bigl(\mathcal U_{\mathcal E,\sigma}(\beta)\otimes \eta\bigr)(\one_S\otimes V^\dagger) \nonumber\\
=&(\id_S\otimes \mathcal P)(\mathcal U_{\mathcal E,\sigma}(\beta)).
\end{align}
Therefore, the corresponding posteriors, the priors for next-round update, also share the same marginal on system $S$. The logic can be repeated and thus after any admissible update sequence, $\beta$ and $\tilde\beta$ produce the same updated belief on $S$. Hence they are iteratively equivalent.
\end{proof}

\section{Candidate-state ensemble beliefs}
\label{sec:candidate-ensemble-belief}

As argued in Section~\ref{sec: formal}, iterative belief updating is applicable to the setting where a source prepares copies of a system in the same unknown state.  More generally, consider an inference task where a source prepares quantum systems, denoted by $S_1,S_2,\ldots$, and sequential experiments are carried out on individual system copies. The sequence is called {\it exchangeable} if changing the order in which the systems are received does not change the global state (permutation invariant) and this symmetry holds for arbitrarily large number of systems. It is a natural belief to adopt if no additional information is available  about the source~\cite{CavesFuchsSchack2002QB,CavesFuchsSchack2002dF}. The quantum de Finetti representation theorem~\cite{Hudson1976,CavesFuchsSchack2002dF} then implies that such a state of $N$ copies has a unique form, 
\begin{equation}\label{eq:finite_exchangeable_prior}
\Omega_{S_1\cdots S_N}=\int p(\rho)\rho^{\otimes N}\,d\rho.
\end{equation}
The agent's ignorance is entirely in the classical probability density function $p(\rho)$ over density operators of a single-copy, and it is this belief that is updated upon receiving new evidence.

It is rather intuitive (and rigorously provable from global source belief in the limit of infinitely-many copies, see Appendix~\ref{app:cse-source}) that the natural assumption of exchangeability leads to the particularly important class of {\it candidate-state beliefs} $\{p(x),\rho_x\}$, where the agent has classical uncertainty over a fixed family of candidate quantum states, as commonly used in Bayesian quantum state tomography~\cite{BQST2010}. Here, $x$ labels the candidate hypothesis, $p(x)$ is the agent's prior probability for that hypothesis, and $\rho_x$ is the corresponding state of system $S$. This class captures the agent's uncertainty about the states prepared by the source, in agreement with the exchangeable source model
\begin{equation}
    \{p(x),\rho_x\}\quad \rightleftharpoons \quad \Omega_{S_1\cdots S_N}=\sum_x p(x)\rho_x^{\otimes N},\quad{N\geq 1}
\end{equation}

In each shot the agent updates $p(x)$ while keeping the family of candidate states fixed. To prevent the update of the candidate states, the suitable one-copy extended belief is
\begin{equation}
\label{eq:candidate-ensemble-belief}
\beta=\sum_x p(x)\,\ketbra{\psi_x}{\psi_x}_{SR_1}\otimes \ketbra{x}{x}_{R_2},
\end{equation}
where $\ket{\psi_x}$ is a purification of $\rho_x$ by the system $R_1$, and $\ket{x}$ are orthonormal states of the classical register $R_2$. We emphasize that the agent has only access to $S$. Indeed, the equation for a single update Eq.~\eqref{eq:extended-update} becomes 
\begin{equation}
    \label{eq:candidate-ensemble-update}
\cU_{\cE,\sigma}(\beta)=
\sum_x \underbrace{p(x)\Tr\left(L\rho_x\right)}_{=p'(x)} \ketbra{\psi_x}{\psi_x}_{SR_1}\otimes \ketbra{x}{x}_{R_2}
\end{equation} where we have denoted
\begin{equation}\label{eq: L candidate state}
L\eqdef 
\cE^{\dagger}\!\left(\cE(\beta_S)^{-1/2}\sigma\,\cE(\beta_S)^{-1/2}\right),\; \beta_S\eqdef\sum_x p(x) \rho_x.
\end{equation} We now state and prove the main result of this paper.

\begin{theorem}[Equivalence condition for candidate-state ensemble beliefs]
\label{thm:proper}
Let
\begin{equation}
    \beta=\sum_x p(x)\,\ketbra{\psi_x}{\psi_x}\otimes \ketbra{x}{x}
\end{equation}
and
\begin{equation}
\gamma=\sum_y q(y)\,\ketbra{\phi_y}{\phi_y}\otimes \ketbra{y}{y}
\end{equation}
be two candidate-state ensemble beliefs, with $\ketbra{\psi_x}{\psi_x}$ and $\ketbra{\phi_y}{\phi_y}$ being any purifications of $\rho_x$ and $\varrho_y$, respectively. Fix $n\ge 1$. Then the following are equivalent:
\begin{enumerate}[label=(\roman*)]
    \item $\beta\sim_n\gamma$.
    \item Their $(n+1)$-th moments are equal:
    \begin{align}
    \sum_x p(x)\,{\rho_x}^{\otimes (n+1)}
    &=
    \sum_y q(y)\,{\varrho_y}^{\otimes (n+1)}.
    \end{align}
\end{enumerate}
\end{theorem}

\begin{proof}
We proceed by induction on $n$.

For $n=1$, the necessary and sufficient condition for general quantum beliefs is known \cite{liu2026properimproper}, i.e.,
\begin{equation}\label{eq:one-shot-equivalence}
    \Tr_{R_\beta R_{\beta}'}\left[\kettbbra{\sqrt{\beta}}\right] = \Tr_{R_{\gamma}R_{\gamma}'}\left[\kettbbra{\sqrt{\gamma}}\right],
\end{equation}
where $\kett{A}$ denotes $\sum_{i,j,i',j'}\bra{i,j}A\ket{i',j'}\ket{i,j,i',j'}_{SRS'R'}$, with systems $S'$ and $R'$ being isomorphic to $S$ and $R$, respectively. Inserting $\beta=\sum_x p(x)\,\ketbra{\psi_x}{\psi_x}\otimes \ketbra{x}{x}$ and $\gamma=\sum_y q(y)\,\ketbra{\phi_y}{\phi_y}\otimes \ketbra{y}{y}$, we obtain
\begin{equation}\label{eq:one-shot-ensemble-equivalence1}
    \sum_x p(x) \rho_x\otimes\rho_x^\tp=\sum_y q(y) \varrho_y\otimes\varrho_y^\tp.
\end{equation}
The calculation is given in Appendix~\ref{app:single-shot-reduction}. Since partial transpose is invertible, Eq.~\eqref{eq:one-shot-ensemble-equivalence1} is equivalent to 
\begin{equation}\label{eq:one-shot-ensemble-equivalence2}
     \sum_x p(x) \rho_x\otimes\rho_x=\sum_y q(y) \varrho_y\otimes\varrho_y.
\end{equation}
Therefore, $\beta\sim_1\gamma$ if and only if their second moments are identical. 

We now assume that the theorem holds for some fixed $n\ge 1$, and prove the equivalence for $n+1$.\\

\paragraph*{Direction $(ii)\implies (i)$. } 
Assume that
\begin{equation}
\sum_x p(x)\, \rho_x^{\otimes (n+2)}
=
\sum_y q(y)\, \varrho_y^{\otimes (n+2)}.
\label{eq:suff-assumption}
\end{equation}
By tracing out any $n+1$ subsystems in Eq.~\eqref{eq:suff-assumption} we obtain 
\begin{equation}
\beta_S=\gamma_S.
\end{equation}
Consider an arbitrary first update $(\cE,\sigma)$. Since the beliefs about reduced density operator of system $S$ coincide, the operator $L$ is common to both beliefs:
\begin{equation}
L\eqdef \cE^{\dagger}\!\left(
\cE(\beta_S)^{-1/2}\sigma\,\cE(\beta_S)^{-1/2}
\right).
\end{equation}
By Eq.~\eqref{eq:candidate-ensemble-update}, the once-updated beliefs are again candidate-state ensembles,
\begin{align}
\cU_{\cE,\sigma}(\beta)
&=
\sum_x p(x)\Tr(\rho_xL)\,\ketbra{\psi_x}{\psi_x}\otimes \ketbra{x}{x},
\\
\cU_{\cE,\sigma}(\gamma)
&=
\sum_y q(y)\Tr(\varrho_yL)\,\ketbra{\phi_y}{\phi_y}\otimes \ketbra{y}{y}.
\end{align}
Their order-$(n+1)$ moments also coincide
    \begin{align}
\sum_x &p(x)\Tr(\rho_xL)\,\rho_x^{\otimes (n+1)}\\
&=
\Tr_1\!\Bigl[(L\otimes \one^{\otimes (n+1)})\sum_x p(x)\, \rho_x^{\otimes (n+2)}\Bigr]
\\
&=
\Tr_1\!\Bigl[(L\otimes \one^{\otimes (n+1)})\sum_y q(y)\, \varrho_y^{\otimes (n+2)}\Bigr] \\
&= \sum_y q(y)\Tr(\varrho_yL)\,\varrho_y^{\otimes (n+1)}
\end{align}
where we used Eq.~\eqref{eq:suff-assumption} to go between the second and the third lines. Hence by the induction hypothesis we have $\cU_{\cE,\sigma}(\beta)\sim_n \cU_{\cE,\sigma}(\gamma)$ for every admissible first update. Finally, this implies  $\beta\sim_{n+1}\gamma$ by virtue of Proposition~\ref{prop:recursive}, proving the first direction.\\

\paragraph*{Direction $(i)\implies (ii)$.} 
Assume now that $\beta\sim_{n+1}\gamma$. By Proposition~\ref{prop:padding}, we have $\beta\sim_1\gamma$, so the initial beliefs about the reduced density operator of system $S$ coincide
\begin{equation}
\beta_S=\gamma_S.
\end{equation}
Let \(\Pi\) project onto \(\supp(\beta_S)\).
Choose some positive semi-definite $E$ satisfying \(E=\Pi E\Pi\) and \(\Tr(\beta_S E)=1\). For the first channel-evidence pair, take
\(\mathcal E=\mathrm{id}\) and
\(\sigma=\sqrt{\beta_S}E\sqrt{\beta_S}\). After the first update, via Eqs.~\eqref{eq:candidate-ensemble-update} and \eqref{eq: L candidate state}, we find $L=E$ and 
\begin{align}
\cU_{\cE,\sigma}(\beta)=
\sum_x p(x)\Tr\left(E\rho_x\right) \ketbra{\psi_x}\otimes \ketbra{x}\\
\cU_{\cE,\sigma}(\gamma)=
\sum_y q(y)\Tr\left(E\varrho_y\right) \ketbra{\phi_y}\otimes \ketbra{y}.
\end{align}
By Proposition~\ref{prop:recursive}, we must have $\cU_{\cE,\sigma}(\beta) \sim_n \cU_{\cE,\sigma}(\gamma)$, which by the induction hypothesis implies equality of the corresponding (n+1)-moments
\begin{align}
\sum_x p(x)\Tr\left(E\rho_x\right) \rho_x^{\otimes (n+1)} = 
\sum_y q(y)\Tr\left(E\varrho_y\right) \varrho_y^{\otimes (n+1)}.
\end{align}
The operators \(E\) span the Hermitian operators on
\(\supp(\beta_S)\). Since all candidate states with nonzero
weight are supported on this subspace, choosing a
tomographically complete set of such operators in the
preceding identity implies
\begin{align}
\sum_x p(x) \, \rho_x\otimes \rho_x^{\otimes (n+1)} = 
\sum_y q(y) \, \varrho_y \otimes \varrho_y^{\otimes (n+1)}
\end{align}
and  completes the proof. 
\end{proof}
For continuous candidate-state ensembles, the discrete
weights are replaced by a probability measure $\nu(dx)$
on the candidate labels, with measurable candidate states
$\rho_x$. The system marginal and update are
$\beta_S=\int\rho_x\,\nu(dx)$ and
$\nu'(dx)=\Tr(L\rho_x)\,\nu(dx)$, respectively, with $L$
defined by Eq.~\eqref{eq: L candidate state}. Theorem~\ref{thm:proper} remains valid with sums replaced by
integrals. Expressions involving continuous classical
registers below are understood as shorthand for this
measure-valued ensemble description.

\subsection{Interpretation of the \texorpdfstring{$(n+1)$}{(n+1)}-th moment condition}
\label{subsec:interpretation}

The update rule used in this article allows for a general forward channel $\mathcal{E}$ and a
general output state $\sigma$ at any iteration. But our result takes on a particularly appealing interpretation if we specialize to the way information about states is usually acquired: one performs a measurement and records a classical outcome. In this subsection we
therefore specialize to update sequences of the form
\begin{equation}
\mathbf{u}
=
\bigl((\cM_1,\ketbra{a_1}{a_1}),\ldots,
(\cM_n,\ketbra{a_n}{a_n})\bigr),
\label{eq:measurement_update_sequence}
\end{equation}
where each forward process is a quantum-to-classical measurement channel
\begin{equation}
\cM_i(\rho)=\sum_a \Tr(E^{(i)}_a\rho)\,\ketbra{a}{a}.
\label{eq:measurement_channel}
\end{equation}
Here $\{E^{(i)}_a\}_a$ is the POVM used in the $i$th round, and the orthonormal
states $\{\ket{a}\}$ encode the classical readout.

Consider a candidate-state ensemble belief
\begin{equation}
\beta^{(0)}=\sum_x p(x)\,\ketbra{\psi_x}{\psi_x}_{SR_1}\otimes \ketbra{x}{x}_{R_2},
\end{equation}
where $\ket{\psi_x}$ is a purification of $\rho_x$. Let
$\beta^{(t-1)}_S=\sum_x p(x|\mathbf{a}_{t-1})\rho_x$ be the system marginal after
$t-1$ measurements. Since $\cM_t(\beta^{(t-1)}_S)$ is diagonal in the outcome
basis, observing $a_t$ gives
\begin{equation}
\begin{split}
    L_t
=&
\cM_t^\dagger\!\left[
\cM_t(\beta^{(t-1)}_S)^{-1/2}
\ketbra{a_t}{a_t}
\cM_t(\beta^{(t-1)}_S)^{-1/2}
\right]
\\=&
\frac{E^{(t)}_{a_t}}
{\sum_{x'}p(x'|\mathbf{a}_{t-1}) \Tr(E^{(t)}_{a_t}\rho_{x'})}.
\end{split}
\end{equation}
Substituting this expression into Eq.~\eqref{eq:candidate-ensemble-update} yields
\begin{equation}
p(x|\mathbf{a}_t)
=
\frac{p(x|\mathbf{a}_{t-1})\Tr(E^{(t)}_{a_t}\rho_x)}
{\sum_{x'}p(x'|\mathbf{a}_{t-1})\Tr(E^{(t)}_{a_t}\rho_{x'})}.
\label{eq:measurement_single_step_bayes}
\end{equation}
Iterating Eq.~\eqref{eq:measurement_single_step_bayes} gives
\begin{equation}
p(x|{\mathbf{a}})
=
\frac{
p(x)\prod_{i=1}^n \Tr(E^{(i)}_{a_i}\rho_x)
}{
\sum_{x'}p(x')\prod_{i=1}^n \Tr(E^{(i)}_{a_i}\rho_{x'})
},
\label{eq:measurement_record_posterior}
\end{equation}
for any outcome record $\mathbf{a}=(a_1,\ldots,a_n)$ with nonzero probability.
The updated belief is therefore
\begin{equation}
\beta^{(n)}(\mathbf{u})
=
\sum_x p(x|{\mathbf{a}})\,
\ketbra{\psi_x}{\psi_x}_{SR_1}\otimes \ketbra{x}{x}_{R_2},
\label{eq:measurement_updated_belief}
\end{equation}
and the corresponding reduced density operator of system $S$ is
\begin{equation}
\Tr_R[\beta^{(n)}(\mathbf{u})]=\sum_x p(x|\mathbf{a})\,\rho_x.
\label{eq:measurement_posterior_system_state}
\end{equation}

This expression can be obtained following the approach of Ref.~\cite{SchackBrunCaves2001}. Take an agent with the candidate-state ensemble belief $\{p(x),\rho_x\}$ and exchangeable source model assumption, who holds $n+1$ copies of the system (the reasoning would be identical for $n+m$). Before the $n$ measurements are performed, the state assigned to the $n+1$ copies is $\Omega_{S_1\cdots S_{n+1}}=\sum_x p(x)\rho_x^{\otimes (n+1)}$. Under the measurements and outcomes of Eq.~\eqref{eq:measurement_update_sequence} on the first $n$ copies, the unnormalized post-measurement state of the
unmeasured $(n+1)$st copy is
\begin{align}
\widetilde{\Omega}_{S_{n+1}|\mathbf{a}}
&=
\Tr_{S_1\cdots S_{n}}\!\left[
\left(E^{(1)}_{a_1}\otimes\cdots\otimes E^{(n)}_{a_n}\otimes\one\right)
\Omega_{S_1\cdots S_{n+1}}
\right]
\nonumber\\
&=
\sum_x p(x)
\prod_{i=1}^n\Tr(E^{(i)}_{a_i}\rho_x)\,\rho_x,
\label{eq:unnormalized_unmeasured_copy}
\end{align}
whence after normalization 
\begin{equation}
\Omega_{S_{n+1}|\mathbf{a}}
=\sum_x p(x|\mathbf{a})\,\rho_x
\label{eq:unmeasured_copy_equals_posterior}
\end{equation} because of Eq.~\eqref{eq:measurement_record_posterior}. So the result of the iterative update Eq.~\eqref{eq:measurement_posterior_system_state} is the same as the reduced density operator of the untouched copy after the
global exchangeable state has been measured and conditioned on the observed
classical data.

This gives the moment condition its operational meaning. Two agents holding two candidate-state ensemble beliefs with the same $(n+1)$-th moments will assign the same exchangeable global state to the $n$ measured
copies and the one untouched copy. Performing measurements and
conditioning on the same outcomes must therefore give the same
post-measurement state. In particular, the reduced density operator of the untouched $(n+1)$-th copy is the
same. Conversely, if the $(n+1)$-copy priors differ, then the difference is, in
principle, a difference in the global state to which the measurement-and-conditioning procedure is applied.

Thus, equality of the $(n+1)$-th moments is not merely
an algebraic criterion. It indicates that the two candidate-state ensemble beliefs represent the
same state assignment for the $n$ systems that will be measured together with
one further system reserved for prediction. Theorem~\ref{thm:proper} extends this
intuition to the general retrodictive framework.

\subsection{Example}
\label{subsec:design-surrogates}

As illustration, let us consider a specific candidate-state ensemble. For a given state $\rho$, the  so-called {\it scrooge ensemble}~\cite{jozsa1994lower} $\{(\mu_\rho(\Psi) \dd\Psi, \ketbra{\Psi_\rho})\}$,
with
\begin{align}
    \mu_\rho(\Psi) &=  \Tr (\ketbra{\Psi} \rho) \quad \text{and} \quad 
    \ket{\Psi_\rho} = \frac{\sqrt{\rho} \ket{\Psi}}{\| \sqrt{\rho }\ket{\Psi}\|},
\end{align}
is obtained from the Haar measure $\dd \Psi$, with the normalizaton choice $\openone = \int \dd \Psi \ketbra{\Psi}$, by left and right multiplication with $\sqrt{\rho}$,
implying 
\begin{equation}
 \int \mu_\rho(\Psi) \dd\Psi \, \ketbra{\Psi_\rho} = \sqrt{\rho} \left(\int \dd\Psi \, \ketbra{\Psi}\right)\sqrt{\rho}  =\rho.   
\end{equation}

 Naturally, this ensemble decomposition gives rise to the continuous candidate-state \emph{scrooge belief}
\begin{align}\label{eq: scrooge}
    \beta = \int \mu_{\rho}(\Psi) \dd \Psi  \ketbra{\Psi_{\rho}}_S \otimes [\Psi]_{R_2}
\end{align}
where $\beta_S=\rho$ and $R_2$ is a classical systems carrying a classical description of the quantum state $\Psi$, a continuous valued random variable. Interestingly, among all pure candidate-state ensemble beliefs of $\rho$ the scrooge ensemble minimizes the accessible information about the preparation label~\cite{jozsa1994lower,fuchs1996distinguishability}, and can thus be chosen as the least informative belief about such preparation procedure in the spirit of Jaynes' principle~\cite{jaynes2003probability}. For the same reason the scrooge ensemble naturally appears as maximally entropic statistical description of quantum states generated by chaotic many-body dynamics or quantum measurements~\cite{mark2024maximum,mok2026nature,liu2026emergence}, i.e. the finite-temperature generalization of Haar randomness.

Analogously to state $t$-designs \cite{Renes2004SIC,ambainis2007quantum} being discrete approximations of the Haar measure with matching lowest moments, it is natural to define the scrooge $t$-designs as follows.

\begin{definition}[\cite{mok2026nature}]
A scrooge $t$-design of a state $\rho$ is an ensemble $\{(p_x, \sigma_x)\}$ which reproduces the moments of the scrooge ensemble up to the order $t$, i.e.
\begin{equation}\label{eq: moment-match}
    \int \mu_\rho(\Psi) \dd \Psi \ketbra{\Psi_\rho}^{\otimes t} = \sum_x p_x \sigma_x^{\otimes t}.
\end{equation}
\end{definition}
\noindent For the special case $\rho =\openone/d$, this is precisely the defining moment condition of a weighted state \(t\)-design, recalling that with our normalization $d^{-1}\mathrm{d}\Psi$ is the Haar probability measure.

By construction, Eq.~\eqref{eq: moment-match} implies the equality of moments for all $k\leq t$, as $\Tr \ketbra{\Psi_\rho}= \Tr \sigma_x=1$, thus any scrooge $(k+1)$-design is also a scrooge  $k$-design. Furthermore, for $t\geq 2$ the lhs of Eq.~\eqref{eq: moment-match} is supported on the totally symmetric subspace of the product Hilbert space, identified by the projector $\Pi_{\rm sym}$, implying the same for the rhs. Hence, any scrooge $t$-design must satisfy $\Tr \left(\Pi_{\rm sym} \sigma_x^{\otimes t}\right) =1$ for all $x$, implying that all states $\sigma_x =\ketbra{\phi_x}$ are pure. Such a scrooge design can be associated with any candidate-state ensemble beliefs
\begin{equation}\label{eq: gamma t}
    \gamma = \sum_x p_x \ketbra{\phi_x}_{S}\otimes \ketbra{x}_{R_2},
\end{equation}
leading to the following immediate Corollary of Theorem~\ref{thm:proper}.

\begin{corollary}\label{cor:design-haar-equivalence}
 A candidate-state ensemble belief $\gamma$ is $n$-shot equivalent to the scrooge belief $\beta$ (Eq.~\ref{eq: scrooge}) with $\rho=\beta_S=\gamma_S$, if and only if it corresponds to a scrooge $(n+1)$-design. If $\rho=\beta_S=\gamma_S=\openone/d$ is maximally mixed, where $d$ is the dimension of the system $S$, this is equivalent to $\gamma$ corresponding to a state $(n+1)$-design.
 \end{corollary}

Corollary~\ref{cor:design-haar-equivalence} gives a concrete computational
application of the moment condition. If an agent conditions on at most
$n$ pieces of evidence, then the continuous scrooge-ensemble distribution over all pure states can be
replaced, without changing any posterior on $S$ for any update sequence of length $n$, by any finite scrooge $(n+1)$-design. The replacement reduces the representation of the prior from a continuous measure to a finite weighted ensemble for a fixed update depth. Thus, bounded-depth quantum Bayesian updates admit exact finite-support surrogate priors. 

\begin{figure}[t]
    \centering
    \includegraphics[width=0.99\linewidth]{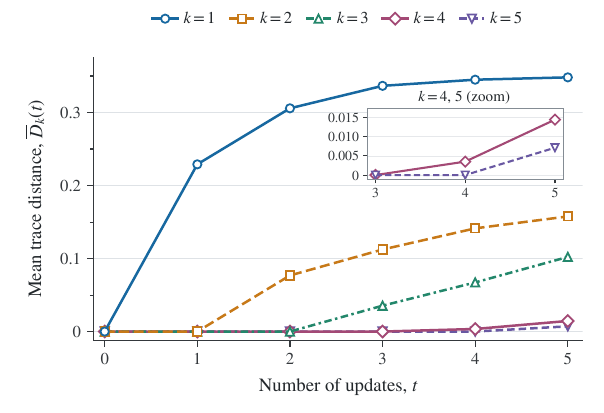}
    \caption{Mean trace distance between design posterior and Haar posterior.
    The plot shows the arithmetic means of trace distance, i.e.,
    $\overline D_k(t)$ in Eq.~\eqref{eq:average-distance-to-haar}, for $k=1,2,3,4,5$. }
    \label{fig:average-design-distance}
\end{figure}

We illustrate this point for a qubit, by comparing the posterior of $k$-design prior with the posterior of Haar prior. A pure qubit state is written as
\begin{equation}
    \rho(\mathbf r)=\frac{1}{2}\left(\one+\mathbf r\cdot\bm\sigma\right),
    \qquad \lvert\mathbf r\rvert=1,
\end{equation}
where $\mathbf r=(x,y,z)$ is a Bloch vector and
$\bm\sigma=(\sigma_x,\sigma_y,\sigma_z)$ is the vector of Pauli matrices. For
qubits, any spherical $k$-design \cite{DELSARTE1991} on the Bloch sphere gives a
projective $k$-design, because every matrix element of
$\rho(\mathbf r)^{\otimes k}$ is a polynomial of degree at most $k$ in the components of $\mathbf r$. Therefore, we can obtain design priors by following the standard theory of spherical designs~\cite{DELSARTE1991,HardinSloane1992New4Designs,hardin1996mclaren}. For each design order $k$, we choose a representative projective design of exact strength $k$, meaning that it is a projective $k$-design but not a projective $(k+1)$-design.  This choice is made to make the hierarchy clear. For explicit state ensemble used in the simulation, refer to Appendix~\ref{app:design-ensemble-choice}.

Consider the same kind of update sequence as in Eq.~\eqref{eq:measurement_update_sequence}, where the measurement is chosen to be the standard qubit SIC POVM~\cite{Renes2004SIC} for the simulation, i.e.,
\begin{equation}
\label{eq:qubit-sic-povm}
    E_a=\frac{1}{4}\left(\one+\mathbf s_a\cdot\bm\sigma\right),
    \qquad a=0,1,2,3,
\end{equation}
where the four SIC vectors $\{\mathbf s_a\}_{a=0}^3$ are $\mathbf s_0=(0,0,1)$ and for $a=1,2,3$,
\[\mathbf s_a=
    \left(
    \frac{2\sqrt2}{3}\cos\frac{2\pi a}{3},
    \frac{2\sqrt2}{3}\sin\frac{2\pi a}{3},
    -\frac13
    \right),
\] 
satisfying $\mathbf s_a\cdot \mathbf s_{a'}=\frac{4\delta_{a a'}-1}{3}$. The measurement yields a sequence of outcomes: $a_1, a_2,\cdots$. Based on the update sequence, one can obtain the updated beliefs starting from projective $k$-design priors and Haar prior, respectively, which we call $k$-design posterior and Haar posterior. We consider design prior up to order 5 and the update sequence of length 5. 
We therefore define
\begin{equation}
\label{eq:average-distance-to-haar}
    \overline D_k(t)=\frac{1}{4^5}\sum_\mathbf{u}
    D\!\left(
        \Tr_R[\beta^{(t)}_{k{\rm -design}}(\mathbf{u})],
        \Tr_R[\beta^{(t)}_{\rm Haar}(\mathbf{u})]
    \right),
\end{equation}
where $D(\rho_1,\rho_2)=\frac{1}{2}\lVert\rho_1-\rho_2\rVert_1$
is the trace distance. For qubit states with Bloch vectors $\mathbf r_1$ and $\mathbf r_2$, this reduces to
\begin{equation}
    D\bigl(\rho(\mathbf r_1),\rho(\mathbf r_2)\bigr)
    =\frac{1}{2}\lvert\mathbf r_1-\mathbf r_2\rvert.
\end{equation}
It is the trace distance of $k$-design posterior and Haar posterior after the first $t$ update averaged over all possible outcome records.

The simulation result is shown in Fig.~\ref{fig:average-design-distance}. It shows that the $k$-design curve stays zero for
$t\le k-1$, exactly as predicted by the moment condition, and can become nonzero at $t=k$. In other words, $(k+1)$-design posteriors agree with Haar posterior up to $k$-th update. This illustrates the operational content of the hierarchy:
for quantum Bayesian update conditioned on record of length at most $n$, Haar prior can be replaced by a finite weighted ensemble that forms a projective $(n+1)$-design.

\section{Conclusion}
We have extended quantum belief equivalence from a single-shot update to iterative, multi-shot updates, with each extended posterior serving
as the prior for the next round. This defines a hierarchy of
equivalence relations that distinguishes beliefs according to their
responses to successively longer evidence sequences. For arbitrary
extended beliefs, we established a recursive characterization of this
hierarchy and showed that reversible transformations of the hidden reference preserve equivalence at every update depth.

The physical interpretation of iterative updating requires a model
relating the systems from which evidence is acquired. An exchangeable
source provides such a model through the quantum de Finetti
representation, motivating candidate-state ensemble beliefs. Within
this class, updating changes the probabilities assigned to fixed
candidate states. Our main result gives a complete characterization:
two candidate-state ensemble beliefs are $n$-shot equivalent if and
only if their $(n+1)$-th moment operators coincide. The moment condition has an operational interpretation through
measurements on an exchangeable source: global state of $n$ measured copies and one
untouched copy accounts for the required moment order $n+1$. It also
shows that finite Scrooge $(n+1)$-designs, including projective designs
in the Haar case, exactly reproduce the corresponding continuous
prior's posterior system states through at most $n$ updates.

Several questions remain open. A closed-form characterization of
multi-shot equivalence for arbitrary extended beliefs is unknown. It would also be interesting to investigate whether a similar moment hierarchy emerges under alternative quantum belief update and retrodiction rules~\cite{Tsa-PRA22,PB23,PF23}. On the practical side, replacing continuous priors with finite design surrogates offers a systematic way to compress prior representations in finite-data Bayesian quantum state tomography~\cite{BQST2010,ABQST2012,Granade_2017}, especially those applying Petz recovery map~\cite{murk2026connectingquantumtomographyquantum,sidajaya2026tomographiclimitspetzrecovery}.
An approximate version of the present result could relate
errors in moment matching to differences between posterior system states. Finally, the connection to Scrooge designs suggests potential applications to statistical ensembles arising in chaotic many-body dynamics and deep thermalization~\cite{mark2024maximum,mok2026nature,liu2026emergence}.

\begin{acknowledgments}
 The authors used OpenAI's ChatGPT (GPT-6 Astra) to assist in checking mathematical arguments and revising the organization and exposition of the manuscript, particularly Appendix C. The authors independently reviewed and verified the final arguments and take full responsibility for the manuscript.
 M.L. thanks Ge Bai, Peter Sidajaya and Clive Cenxin Aw for helpful discussions. This project is supported by the National Research Foundation, Singapore through the National Quantum Office, hosted in A*STAR, under its Centre for Quantum Technologies Funding Initiative (S24Q2d0009). V.S.~acknowledges support from the Ministry of Education, Singapore, under the Tier 2 grant ``Bayesian approach to irreversibility'' (Grant No.~MOE-T2EP50123-0002).
\end{acknowledgments}

\bibliography{refs}

\appendix\newpage
\onecolumngrid

\section{Classical limit of the prior-extended Petz update}
\label{app:classical-limit}
We show that the prior-extended Petz update reduces to Jeffrey's rule
when the belief, channel, and output state are classical. Consider a
general classical belief on $S$ and $R$,
\begin{equation}
    \beta_{SR}
    =
    \sum_{x,r} p(x,r)
    \ketbra{x}_S\otimes\ketbra{r}_R.
\end{equation}
Its marginal on $S$ is
\begin{equation}
    \beta_S
    =
    \Tr_R[\beta_{SR}]
    =
    \sum_x p(x)\ketbra{x}_S,
    \qquad
    p(x)=\sum_r p(x,r).
\end{equation}
Let the channel $\cE:S\to T$ be classical in the chosen bases,
\begin{equation}
    \cE\bigl(\ketbra{x}_S\bigr)
    =
    \sum_y \mathcal{P}(y|x)\ketbra{y}_T,
\end{equation}
and let the evidence state be
\begin{equation}
    \sigma
    =
    \sum_y q(y)\ketbra{y}_T.
\end{equation}
The output distribution
predicted by the prior is
\begin{equation}
    \mathcal{P}_*p(y)
    =
    \sum_x p(x)\mathcal{P}(y|x),
\end{equation}
and hence
\begin{equation}
    \cE(\beta_S)
    =
    \sum_y \mathcal{P}_*p(y)\ketbra{y}_T.
\end{equation}
We assume that $q(y)>0$ only if $\mathcal{P}_*p(y)>0$, so that all
inverses below are defined on the relevant support.

Since the states and channel are classical,
\begin{equation}
    \cE(\beta_S)^{-1/2}
    \sigma
    \cE(\beta_S)^{-1/2}
    =
    \sum_y
    \frac{q(y)}{\mathcal{P}_*p(y)}
    \ketbra{y}_T.
\end{equation}
Moreover,
\begin{equation}
    \cE^\dagger\bigl(\ketbra{y}_T\bigr)
    =
    \sum_x \mathcal{P}(y|x)\ketbra{x}_S,
\end{equation}
which gives
\begin{equation}
    \cE^\dagger\left(
        \cE(\beta_S)^{-1/2}
        \sigma
        \cE(\beta_S)^{-1/2}
    \right)
    =
    \sum_x
    \left[
        \sum_y
        \frac{q(y)\mathcal{P}(y|x)}
             {\mathcal{P}_*p(y)}
    \right]
    \ketbra{x}_S.
\end{equation}
Substituting this expression into Eq.~\eqref{eq:extended-update} yields
\begin{equation}
    \cU_{\cE,\sigma}(\beta)
    =
    \sum_{x,r}
    p(x,r)
    \left[
        \sum_y
        \frac{q(y)\mathcal{P}(y|x)}
             {\mathcal{P}_*p(y)}
    \right]
    \ketbra{x}_S\otimes\ketbra{r}_R.
\end{equation}
After tracing out the reference system, we obtain
\begin{align}
    \Tr_R\!\left[
        \cU_{\cE,\sigma}(\beta_{SR})
    \right]
    \nonumber=
    \sum_x p'(x)\ketbra{x}_S,
\end{align}
where
\begin{equation}
    p'(x)
    =
    p(x)
    \sum_y
    \frac{q(y)\mathcal{P}(y|x)}
         {\mathcal{P}_*p(y)}.
\end{equation}
Equivalently,
\begin{equation}
    p'
    =
    \mathcal{J}_{\mathcal{P},q}(p),
\end{equation}
which is precisely Jeffrey's rule in Eq.~\eqref{eq: Jeffrey}. Therefore, the prior-extended Petz update recovers Jeffrey's probability-kinematics update in the classical
limit.

\section{Derivation of Eq.~\eqref{eq:one-shot-ensemble-equivalence1} from Eq.~\eqref{eq:one-shot-equivalence}}
\label{app:single-shot-reduction}
We first evaluate the left-hand side of Eq.~\eqref{eq:one-shot-equivalence}. To do so, we expand it in a fixed orthonormal product basis. In particular, the indices $i,l$ and $i',l'$ label the basis of $\Hil_S$ and $\Hil_{S'}$, respectively. The indices $j,m$ and $j',m'$ label the basis of first reference system, while the indices $k,n$ and $k',n'$ correspond to the second reference system, which carries the classical register. With respect to this basis, we obtain
\begin{equation}
\begin{split}
&\Tr_{R_{\beta}R_{\beta}'}\left[\sum_{\substack{x,x',\\i,j,k,i',j',k',\\l,m,n,l',m',n'}}\sqrt{p(x) p(x')}\braket{ij}{\psi_x}\braket{\psi_x}{i'j'}\braket{k}{x}\braket{x}{k'}\left(\braket{lm}{\psi_{x'}}\braket{\psi_{x'}}{l'm'}\braket{n}{x'}\braket{x'}{n'}\right)^*\ketbra{ijki'j'k'}{lmnl'm'n'}\right]\\
=&\Tr_{R_{\beta}R_{\beta}'}\left[\sum_{\substack{x,x'\\i,j,i',j',\\l,m,l',m'}}\sqrt{p(x) p(x')}\braket{ij}{\psi_x}\braket{\psi_x}{i'j'}\left(\braket{lm}{\psi_{x'}}\braket{\psi_{x'}}{l'm'}\right)^*\ketbra{ijxi'j'x}{lmx'l'm'x'}\right]\\
=&\sum_{\substack{x,x'\\i,j,i',j',\\l,m,l',m'}}\sqrt{p(x) p(x')}\braket{ij}{\psi_x}\braket{\psi_x}{i'j'}\braket{l'm'}{\psi_{x'}}\braket{\psi_{x'}}{lm}\delta_{jm} \delta_{j'm'}\delta_{xx'}\ketbra{ii'}{ll'}\\
=&\sum_{\substack{x,i,j,l,i',j',l'}}p(x) \braket{ij}{\psi_x}\braket{\psi_x}{i'j'}\braket{l'j'}{\psi_x}\braket{\psi_x}{lj}\ketbra{ii'}{ll'}=\sum_x p(x) \rho_x\otimes\rho_x^\tp,
\end{split}
\end{equation}
where the transpose is taken with respect to the chosen basis. In the third line, the Kronecker deltas arise from tracing over the reference systems, and $\delta_{xx'}$ sets $x'=x$, so that $\sqrt{p(x)p(x')}=p(x)$. The final equality follows by identifying the remaining coefficients with the matrix elements of $\rho_x\otimes\rho_x^\tp$.

Analogously, the right-hand side of Eq.~\eqref{eq:one-shot-equivalence} is equal to $\sum_y q(y) \varrho_y\otimes\varrho_y^\tp$. Therefore, we have
\begin{equation}
    \sum_x p(x) \rho_x\otimes\rho_x^\tp=\sum_y q(y) \varrho_y\otimes\varrho_y^\tp
\end{equation}

\section{Exchangeable sources and candidate-state ensemble beliefs}
\label{app:cse-source}
The main text encodes uncertainty about an exchangeable source by
candidate-state ensemble belief. Here we justify
this representation from global updates on exchangeable source beliefs.

An exchangeable source emits infinitely many copies of an unknown prototype state. Let $Q_i$ be the auxiliary system associated with $S_i$, and set
$X_i=S_iQ_i$, with $S$ and $Q$ finite-dimensional. A consistent permutation-invariant global belief at a finite cutoff $N$ is in the de Finetti form
\begin{equation}
\Omega_N=\Omega_{X_1\cdots X_N}
=\int\Gamma^{\otimes N}\,\mu(d\Gamma),
\qquad N=1,2,\ldots .
\label{eq:app-definetti-extended}
\end{equation}
Here \(\mu\) is a probability measure over states of \(X=SQ\).
The marginal on \(S_1\cdots S_N\) is the source state considered
in the main text. We first state a technical observation, with its proof postponed to the end.

\begin{lemma}[Asymptotic form of the global update]
\label{lem:appc-asymptotic-update}
For any fixed positive operator \(\Lambda\) on \(X\),
\begin{equation}
    \lim_{N\to\infty}
    \left\|
        \sqrt{\Omega_N}
        \bigl(\Lambda\otimes \one_{X_2\cdots X_N}\bigr)
        \sqrt{\Omega_N}
        -
        \int
        \sqrt{\Gamma}\Lambda\sqrt{\Gamma}
        \otimes\Gamma^{\otimes(N-1)}
        \,\mu(d\Gamma)
    \right\|_1
    =0.
    \label{eq:appc-asymptotic-update}
\end{equation}
\end{lemma}

For an admissible channel--evidence pair \((\mathcal E,\sigma)\)
acting on \(S_j\), take \(\Lambda=L\otimes \one_Q\), where
\begin{equation}
        L=\mathcal E^\dagger\!\left[
            \mathcal E(\bar\gamma)^{-1/2}
            \sigma
            \mathcal E(\bar\gamma)^{-1/2}
        \right],\qquad
        \bar\gamma
        =\int \Gamma_S\,\mu(d\Gamma),
        \qquad
        \Gamma_S=\operatorname{Tr}_Q\Gamma.
    \label{eq:appc-evidence}
\end{equation}
All inverses are taken on the relevant support. Because an update on any copy $S_j$
is related to an update on $S_1$ by relabelling the extended copies, \textit{w.l.o.g.}, we label the updated
copy as copy $1$. The global retrodictive
posterior
\begin{equation}
\Omega^{L}_{N}
=\sqrt{\Omega_N}
((L_{S_1}\otimes\one_{Q_1})\otimes\one_{X_2\cdots X_N})
\sqrt{\Omega_N}
\label{eq:app-normalized-full-posterior}
\end{equation}
is normalized, since admissibility gives
\(\operatorname{Tr}(L\bar\gamma)=1\).

We impose the following global consistency requirement. For every admissible channel--evidence pair, there must exist a
probability measure \(\mu_L\), such that
\begin{equation}
    \lim_{N\to\infty}
    \left\|
        \Omega_N^L
        -
        \int \Delta^{\otimes N}\,\mu_L(d\Delta)
    \right\|_1
    =0.
    \label{eq:appc-consistency}
\end{equation}
This is a consistency requirement on the retrodictive belief
about the source. Neither the support nor the weights of
\(\mu_L\) are prescribed.

\begin{theorem}[Stability of the source under retrodiction]
\label{thm:appc-stability}
Equation~\eqref{eq:appc-consistency} holds for every admissible
channel--evidence pair if and only if, on a common set of full
\(\mu\)-measure,
\begin{equation}
    \sqrt{\Gamma}(L\otimes \one_Q)\sqrt{\Gamma}
    =
    \operatorname{Tr}(L\Gamma_S)\Gamma
    \label{eq:app-stability-condition}
\end{equation}
for every \(L\) obtained from such a pair. The posterior measure
is then uniquely determined by
\begin{equation}
    \mu_L(d\Gamma)
    =
    \operatorname{Tr}(L\Gamma_S)\,\mu(d\Gamma).
    \label{eq:appc-reweighting}
\end{equation}
\end{theorem}
The condition~\eqref{eq:app-stability-condition} yields a non-empty set, because it is manifestly satisfied by every pure \(\Gamma\). In fact, it can be shown that every $\Gamma$ satisfying this condition is, up to
an isometry acting only on the auxiliary system, a pure state
tensored with a junk state.

\begin{definition}[Exchangeable source belief]
\label{def:app-candidate-source-belief}
An exchangeable source belief is a belief about its infinitely many copies satisfying the exchangeable condition
\eqref{eq:app-definetti-extended} and the stability
condition~\eqref{eq:app-stability-condition}.
\end{definition}
Thus updating an exchangeable source belief changes only the component weights. 

For a finite candidate measure
\(\mu=\sum_x p(x)\delta_{\Gamma_x}\) satisfying
Eq.~\eqref{eq:app-stability-condition}, set
\(\rho_x=(\Gamma_x)_S\) and choose a purification
\(\lvert\psi_x\rangle\) of each \(\rho_x\). The candidate-state
ensemble belief
\begin{equation}
    \beta^{\mathrm{cse}}
    =
    \sum_x p(x)
    \lvert\psi_x\rangle\langle\psi_x\rvert_{SR_1}
    \otimes
    \lvert x\rangle\langle x\rvert_{R_2}
    \label{eq:appc-cse}
\end{equation}
has the same system marginal \(\bar\gamma\). Its update therefore
uses the same \(L\) and, by the candidate-state ensemble update
rule in the main text, gives
\begin{equation}
    p_L(x)=p(x)\operatorname{Tr}(L\rho_x),
    \qquad
    \rho_L=\sum_x p_L(x)\rho_x.
    \label{eq:appc-cse-update}
\end{equation}
The infinite-copy global
update and the update of $\beta^{\mathrm{cse}}$ therefore give the same posterior candidate weights and system state after one update. Reweighting preserves the components satisfying
Eq.~\eqref{eq:app-stability-condition}, so the same source update rule continues to apply. Consequently,
the one-step equality repeats through every finite admissible sequence.\\

\begin{remark}
Intuitively, the equivalence between $\beta^{\mathrm{cse}}$ in Eq.~\eqref{eq:appc-cse} and the exchangeable source belief is established immediately by identifying 
an asymptotic number of copies of the system $SQ$ with the classical label $R_2$, i.e., $\lim_{N\to \infty }\bigotimes_{k=2}^N {\Gamma_x}_{S_kQ_k} \simeq \ketbra{x}_{R_2}$. Under the stability condition~\eqref{eq:app-stability-condition}, Eq.~\eqref{eq:appc-consistency} ensures that the global retrodictive posterior remains exchangeable in the infinite-copy limit. Consequently, the one-copy extended posterior also describes the belief about the next copy and can serve as its prior. This justifies iterating the update of $\beta^{\mathrm{cse}}$. If instead a system is discarded after its update, the stability condition in Eq.~\eqref{eq:app-stability-condition} becomes obsolete (it is always true upon applying the trace on both sides), and Eq.~\eqref{eq:appc-consistency} automatically holds for {\it the remaining systems} by Lemma~\ref{lem:appc-asymptotic-update}. Strictly speaking, this is not "retrodiction", as we discard the system for which evidence was obtained.
\end{remark}

\subsection*{Proofs}

\begin{proof}[Proof of Lemma~\ref{lem:appc-asymptotic-update}]
Set \(m=N-1\) and write
\begin{equation}
    K_\Lambda(\Gamma)
    =\sqrt{\Gamma}\Lambda\sqrt{\Gamma},
    \qquad
    \Lambda^{(1)}
    =\Lambda\otimes \one_X^{\otimes m}.
    \label{eq:appc-lemma-notation}
\end{equation}
For any \(\delta>0\), choose a finite Borel partition
\(\{C_i\}\) of the compact state space into cells of trace-norm
diameter at most \(\delta\), with \(\mu(\partial C_i)=0\).
Such a partition can be obtained from a finite cover by
sufficiently small balls with zero-measure boundaries.
Discard cells of zero measure and define
\begin{equation}
    p_i=\mu(C_i),
    \qquad
    \Gamma_i\in C_i,
    \qquad
    \tau_{i,m}
    =
    \frac{1}{p_i}
    \int_{C_i}\Gamma^{\otimes m}\,\mu(d\Gamma).
    \label{eq:appc-partition}
\end{equation}
Approximating only the first copy gives
\begin{equation}
    \widehat\Omega_N
    =
    \sum_i p_i\Gamma_i\otimes\tau_{i,m},
    \qquad
    \|\Omega_N-\widehat\Omega_N\|_1\leq\delta.
    \label{eq:appc-prior-approximation}
\end{equation}
The bound is uniform in \(m\).

For distinct cells, the conditional tail states become
asymptotically distinguishable:
\begin{equation}
    f(\tau_{i,m},\tau_{j,m})\longrightarrow0
    \quad(i\neq j),
    \qquad
    f(\rho,\sigma)=\|\sqrt{\rho}\sqrt{\sigma}\|_1.
    \label{eq:appc-tail-fidelity}
\end{equation}
Indeed, measurements of a fixed informationally complete POVM
on the tail copies give a consistent estimator of \(\Gamma\).
Outside the cell boundaries, this identifies the cell
containing \(\Gamma\) with probability tending to one.
Since the boundaries have zero measure, dominated convergence
gives vanishing classification error conditional on each cell.
Monotonicity of fidelity under this measurement and classical
processing then implies Eq.~\eqref{eq:appc-tail-fidelity}.

Define the positive operator
\begin{equation}
    B_m
    =
    \sum_i\sqrt{p_i}\,
    \sqrt{\Gamma_i}\otimes\sqrt{\tau_{i,m}}.
    \label{eq:appc-root-approximation}
\end{equation}
The Powers--St{\o}rmer inequality and
Eq.~\eqref{eq:appc-tail-fidelity} give
\begin{equation}
    \begin{aligned}
        \|\sqrt{\widehat\Omega_N}-B_m\|_2^2
        &\leq
        \|\widehat\Omega_N-B_m^2\|_1
        \\
        &\leq
        \sum_{i\neq j}
        \sqrt{p_ip_j}\,
        f(\tau_{i,m},\tau_{j,m})
        \longrightarrow0.
    \end{aligned}
    \label{eq:appc-root-bound}
\end{equation}
Here we used \(f(\Gamma_i,\Gamma_j)\leq1\).
Moreover, \(\|B_m\|_2\to1\). The Hilbert--Schmidt
Cauchy--Schwarz inequality therefore implies that
\(\sqrt{\widehat\Omega_N}\Lambda^{(1)}
\sqrt{\widehat\Omega_N}-B_m\Lambda^{(1)}B_m\)
tends to zero in trace norm.

The off-diagonal terms in \(B_m\Lambda^{(1)}B_m\) obey
\begin{equation}
    \begin{aligned}
        &\left\|
            \sqrt{\Gamma_i}\Lambda\sqrt{\Gamma_j}
            \otimes
            \sqrt{\tau_{i,m}}\sqrt{\tau_{j,m}}
        \right\|_1
        \\
        &\qquad\leq
        \|\Lambda\|_\infty
        f(\tau_{i,m},\tau_{j,m})
        \longrightarrow0.
    \end{aligned}
    \label{eq:appc-off-diagonal}
\end{equation}
Thus, for the fixed partition,
\begin{equation}
    \left\|
        \sqrt{\widehat\Omega_N}\Lambda^{(1)}
        \sqrt{\widehat\Omega_N}
        -
        \sum_i p_iK_\Lambda(\Gamma_i)\otimes\tau_{i,m}
    \right\|_1
    \longrightarrow0.
    \label{eq:appc-partition-update}
\end{equation}

It remains to control the approximation uniformly in \(m\).
For normalized states \(\rho,\sigma\) and a positive operator
\(T\), the Powers--St{\o}rmer and Cauchy--Schwarz inequalities
give
\begin{equation}
    \|\sqrt{\rho}T\sqrt{\rho}
      -\sqrt{\sigma}T\sqrt{\sigma}\|_1
    \leq
    2\|T\|_\infty\sqrt{\|\rho-\sigma\|_1}.
    \label{eq:appc-update-continuity}
\end{equation}
Together with Eq.~\eqref{eq:appc-prior-approximation}, this
bounds the change caused by replacing \(\Omega_N\) with
\(\widehat\Omega_N\) by
\(2\|\Lambda\|_\infty\sqrt{\delta}\).
Applying the same inequality within each cell also gives
\begin{equation}
    \begin{aligned}
        &\left\|
            \sum_i p_iK_\Lambda(\Gamma_i)\otimes\tau_{i,m}
            -
            \int
            K_\Lambda(\Gamma)\otimes\Gamma^{\otimes m}
            \,\mu(d\Gamma)
        \right\|_1
        \\
        &\qquad\leq
        2\|\Lambda\|_\infty\sqrt{\delta}.
    \end{aligned}
    \label{eq:appc-component-approximation}
\end{equation}
Combining these bounds with
Eq.~\eqref{eq:appc-partition-update}, we obtain
\begin{equation}
    \begin{aligned}
        &\limsup_{N\to\infty}
        \Bigg\|
            \sqrt{\Omega_N}\Lambda^{(1)}\sqrt{\Omega_N}
            -
            \int
            K_\Lambda(\Gamma)\otimes\Gamma^{\otimes(N-1)}
            \,\mu(d\Gamma)
        \Bigg\|_1
        \\
        &\qquad\leq
        4\|\Lambda\|_\infty\sqrt{\delta}.
    \end{aligned}
    \label{eq:appc-final-approximation}
\end{equation}
The partition is fixed before taking \(N\to\infty\).
Letting \(\delta\downarrow0\) proves the lemma.
\end{proof}

\begin{proof}[Proof of Theorem~\ref{thm:appc-stability}]
Write
\begin{equation}
    K_L(\Gamma)
    =
    \sqrt{\Gamma}(L\otimes \one_Q)\sqrt{\Gamma}.
    \label{eq:appc-local-component-update}
\end{equation}

For sufficiency, assume
Eq.~\eqref{eq:app-stability-condition}.
Substitution into Lemma~\ref{lem:appc-asymptotic-update}
gives Eq.~\eqref{eq:appc-consistency} with the measure in
Eq.~\eqref{eq:appc-reweighting}.
This measure is normalized because
\begin{equation}
    \int
    \operatorname{Tr}(L\Gamma_S)\,\mu(d\Gamma)
    =
    \operatorname{Tr}(L\bar\gamma)
    =1.
    \label{eq:appc-posterior-normalization}
\end{equation}

For necessity, fix an admissible channel--evidence pair and its
operator \(L\), and assume Eq.~\eqref{eq:appc-consistency}.
Contractivity under partial trace, applied to
Eq.~\eqref{eq:appc-consistency} and
Lemma~\ref{lem:appc-asymptotic-update}, yields, for every
fixed \(n\geq1\),
\begin{equation}
    \begin{aligned}
        \lim_{N\to\infty}
        \operatorname{Tr}_{X_{n+1}\cdots X_N}\Omega_N^L
        &=
        \int
        K_L(\Gamma)\otimes\Gamma^{\otimes(n-1)}
        \,\mu(d\Gamma)
        \\
        &=
        \int
        \Delta^{\otimes n}\,\mu_L(d\Delta).
    \end{aligned}
    \label{eq:appc-retained-marginals}
\end{equation}
Tracing out the first copy gives
\begin{equation}
    \int
    \Delta^{\otimes m}\,\mu_L(d\Delta)
    =
    \int
    \operatorname{Tr}(L\Gamma_S)
    \Gamma^{\otimes m}\,\mu(d\Gamma),
    \qquad m\geq0.
    \label{eq:appc-tail-moments}
\end{equation}
Both sides define normalized exchangeable families.
Uniqueness of the de Finetti measure therefore establishes
Eq.~\eqref{eq:appc-reweighting}.

Substituting Eq.~\eqref{eq:appc-reweighting} into
Eq.~\eqref{eq:appc-retained-marginals}, define
\begin{equation}
    M_L(\Gamma)
    =
    K_L(\Gamma)-\operatorname{Tr}(L\Gamma_S)\Gamma.
    \label{eq:appc-component-difference}
\end{equation}
Then
\begin{equation}
    \int
    M_L(\Gamma)\otimes\Gamma^{\otimes m}
    \,\mu(d\Gamma)
    =0,
    \qquad m\geq0.
    \label{eq:appc-vanishing-moments}
\end{equation}
For Hermitian operators \(C\) on \(X\) and \(B\) on
\(X^{\otimes m}\), this implies
\begin{equation}
    \int
    \operatorname{Tr}[CM_L(\Gamma)]\,
    \operatorname{Tr}[B\Gamma^{\otimes m}]
    \,\mu(d\Gamma)
    =0.
    \label{eq:appc-scalar-moments}
\end{equation}
The real linear span of the functions
\(\Gamma\mapsto\operatorname{Tr}[B\Gamma^{\otimes m}]\),
over all \(m\) and Hermitian \(B\), is an algebra containing
the constants and separating points of the compact state
space. By the Stone--Weierstrass theorem, it is dense in the
continuous real-valued functions.

For fixed \(C\), the function
\(f_C(\Gamma)=\operatorname{Tr}[CM_L(\Gamma)]\)
is continuous and real. Approximating \(f_C\) by these
polynomials in Eq.~\eqref{eq:appc-scalar-moments} gives
\begin{equation}
    \int f_C(\Gamma)^2\,\mu(d\Gamma)=0.
    \label{eq:appc-vanishing-square}
\end{equation}
Taking a finite Hermitian basis for \(C\) shows that
\(M_L(\Gamma)=0\) on a set of full \(\mu\)-measure.

To obtain one common full-measure set for all admissible
updates, restrict \(S\) to \(\supp\bar\gamma\).
Positivity ensures that \(\Gamma\) is supported on this
subspace tensored with \(Q\), for \(\mu\)-almost every
\(\Gamma\). On this support, every positive \(L\) satisfying
\(\operatorname{Tr}(L\bar\gamma)=1\) arises from an admissible
pair: choose
\begin{equation}
    \mathcal E=\mathrm{id},
    \qquad
    \sigma=\sqrt{\bar\gamma}\,L\sqrt{\bar\gamma}.
    \label{eq:appc-admissible-realization}
\end{equation}
Choose a finite set of such positive operators spanning the
Hermitian operators on \(\supp\bar\gamma\).
Intersecting their full-measure sets and using linearity in
\(L\) establishes Eq.~\eqref{eq:app-stability-condition}
on a common full-measure set. Operators on the original
system space enter only through their compression to this
support.

The same linearity extends the component identity to every
positive local operator on that support. Reweighting the
measure therefore preserves the identity, including when
the system support shrinks after an update. This justifies
its repeated use with the posterior as the next prior.
\end{proof}

\section{State ensemble for each design in the simulation}
\label{app:design-ensemble-choice}
We list the state ensembles used as design priors in the simulation in terms of their Bloch-sphere coordinates. All points are of equal weight. 
The $1$-design consists of the eigenbasis of Pauli-X, 
\begin{equation}
    \{(1,0,0),(-1,0,0)\}.
\end{equation}
The $2$-design is the tetrahedral set,
\begin{equation}
\label{eq:tetrahedral-design-vectors}
    \{\frac{1}{\sqrt 3}(1,1,1),
    \frac{1}{\sqrt 3}(1,-1,-1),
    \frac{1}{\sqrt 3}(-1,1,-1),
    \frac{1}{\sqrt 3}(-1,-1,1)\}.
\end{equation}
The $3$-design is the octahedral set,
\begin{equation}
\label{eq:octahedral-design-vectors}
    \{(1,0,0), (-1,0,0), (0,1,0), (0,-1,0), (0,0,1), (0,0,-1)\}.
\end{equation}
Let $z_0=\frac{1}{\sqrt5}, r_0=\frac{2}{\sqrt5}$. A twisted-icosahedron point set is
\begin{equation}
\label{eq:twisted-icosahedron-design}
\begin{split}
X_\theta
=
&\ \{(0,0,1),(0,0,-1)\} \cup
\left\{
\left(
r_0\cos\!\left(\frac{2\pi j}{5}+\theta\right),
r_0\sin\!\left(\frac{2\pi j}{5}+\theta\right),
z_0
\right)
\right\}_{j=0}^{4} \\
&\cup
\left\{
\left(
r_0\cos\!\left(\frac{2\pi j}{5}+\frac{\pi}{5}\right),
r_0\sin\!\left(\frac{2\pi j}{5}+\frac{\pi}{5}\right),
-z_0
\right)
\right\}_{j=0}^{4}.
\end{split}
\end{equation}
For the $4$-design, we take the point set $X_{\pi/5}$. The $5$-design is the regular icosahedron, obtained
from the same construction by taking $\theta=0$, i.e., the point set $X_0$.

\end{document}